\documentclass[twocolumn,secnumarabic,amssymb, nobibnotes, aps, prd]{revtex4-1}

\usepackage{graphicx}
\usepackage{amsmath,amsfonts,amssymb,bm}
\usepackage{epstopdf}

\newcommand{\eps}{\varepsilon}	        
\newcommand{\dd}{\partial}

\begin{document}

\title{The characteristic shape of emission profiles of plasma spokes in HiPIMS: the role of secondary electrons}%

\author{A. Hecimovic$^{1}$, S. Gallian$^2$, R.-P. Brinkmann$^2$, M. B\"{o}ke$^1$, J. Winter$^1$}%
\affiliation{$^1$Institute of Experimental Physics II, Ruhr-University Bochum, Germany}
\affiliation{$^2$Institute for Theoretical Electrical Engineering, Ruhr-University Bochum, Germany}
\date{\today}%

\begin{abstract}
A time resolved analysis of the emission of HiPIMS plasmas reveals inhomogeneities in the form of rotating spokes. The shape of these spokes is very characteristic depending on the target material. The localized enhanced light emission has been correlated with the ion production. Based on these data, the peculiar shape of the emission profiles can be explained by the localized generation of secondary electrons, resulting in an energetic electron pressure exceeding the magnetic pressure. This general picture is able to explain the observed emission profile for different target materials including gas rarefaction and second ionization potential of the sputtered elements.
\end{abstract}

\maketitle

\begin{figure}[h!] 
\centering  {\includegraphics[width=200pt]{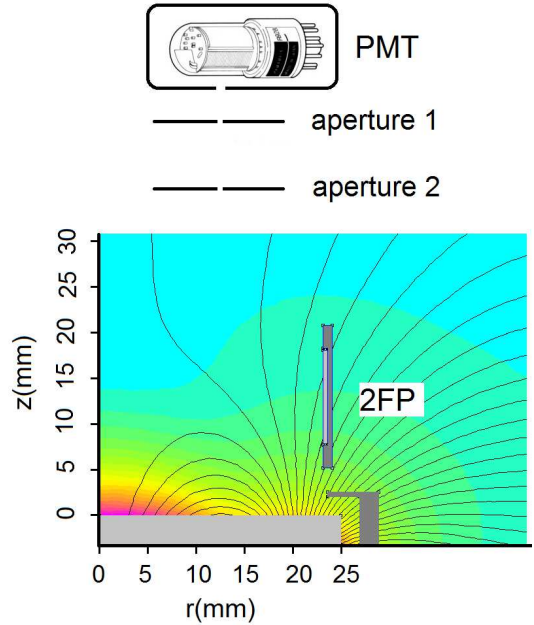}} \caption{Experimental setup showing position of the double flat probe (2FP) and photomultiplier tube (PMT) with 2 apertures }\label{fig1}
\end{figure}

\begin{figure}[h!] 
\centering  {\includegraphics[width=240pt]{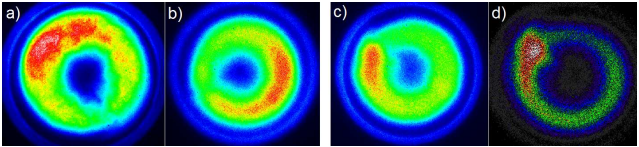}} \caption{Light emission of the plasma for different discharge with a) Ti (13.6 eV), b) Nb (14.0 eV), c) Cr (16.5 eV) and d) Al (18.8 eV) target. The values in the brackets are the second ionisation potential}\label{fig2}
\end{figure}

\begin{figure}[h!] 
\centering  {\includegraphics[width=240pt]{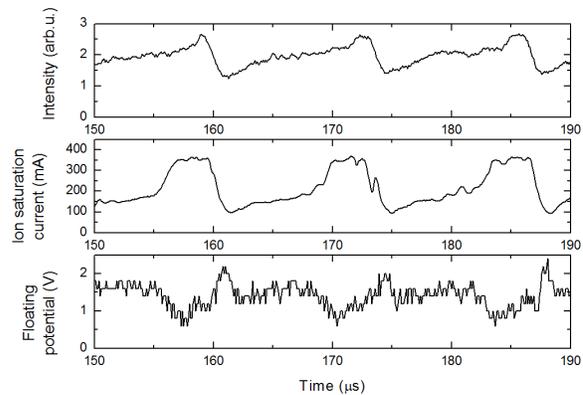}} \caption{Time comparison of the light signal (top), ion saturation current (middle) and floating potential (bottom) oscillations for discharge with Al target}\label{fig3}
\end{figure}

\begin{figure}[h!] 
\centering  {\includegraphics[width=240pt]{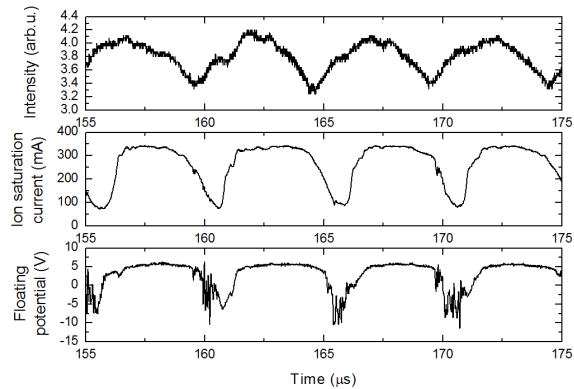}} \caption{Time comparison of the light signal (top), ion saturation current (middle) and floating potential (bottom) oscillations for discharge with Ti target}\label{fig4}
\end{figure}

Recently we have reported on HiPIMS discharges exhibiting spokes as localized light emission rotating in the $\mathbf{E} \times \mathbf{B}$ direction with angular frequencies in the range of a 100 kHz with different mode numbers and with transition from stochastic to periodic behaviour depending on the discharge parameters \cite{ehiasarian_high_2012}, \cite{winter_instabilities_2013}. Kozyrev et al. \cite{kozyrev_optical_2011} and Anders et al.\cite{anders_drifting_2012} have independently observed the same phenomena. Kozyrev et al. explained the inhomogeneity of the discharge as an effect of a high density ion current driven along the $E_\phi \mathbf{e}_\phi \times \mathbf{B}$ drift direction generated by an azimuthal electric field $E_\phi$ \cite{kozyrev_optical_2011}. The azimuthal electric field was attributed to the azimuthal modulation in the plasma density. Anders et al. presented a model where the spoke is an ionization zone where electrons are decelerated due to the interaction with charged particles \cite{anders_drifting_2012}, \cite{anders_self-organization_2012}. The model predicts the existence of an azimuthal electric field due to a reduced electron azimuthal velocity. Similarly to Kozyrev's model, the charged particles are removed from the target due to the $E_\phi \mathbf{e}_\phi \times \mathbf{B}$ drift. Brenning et al. \cite{brenning_alfvens_2012}, \cite{brenning_spokes_2013} offered a physical model to describe the spoke based on the Alfven’s critical ionization velocity (CIV). The ionization within the structure creates an electric field as a result of a charge separation based on the CIV theory. The model postulates a modified two stream instability accelerating the electrons towards the target and generating a drift opposite to the  $\mathbf{E} \times \mathbf{B}$ direction, therefore explaining the rotation speed being one order of magnitude lower then  $\mathbf{E} \times \mathbf{B}$ speed. These phenomenological models can explain the observed lower than  $\mathbf{E} \times \mathbf{B}$ speed of the spokes and the transport of charged particles in the $E_\phi \mathbf{e}_\phi \times \mathbf{B}$ direction. However, these models do not explain the different shapes of the spokes depending on the target material \cite{ehiasarian_high_2012}\cite{anders_drifting_2012}, and the appearance of the ``jets" reported by Ni et al. \cite{ni_plasma_2012}. \\
Here, we measure the locally emitted light and electron and ion saturation current indicating localization of the ionization processes and thereby also of the discharge current. A localized discharge current leads to a localized Ar gas rarefaction, which is responsible for the observed emission shape, and a high density of energetic secondary electrons.  
The experimental setup consists of a double flat probe (2FP) and a photomultiplier tube (PMT) with 2 apertures, as shown in Figure \ref{fig1}. The double flat probe consists of two flat concentric surfaces, one rectangular with a surface of 0.5 cm$^2$ and a surrounding one with surface of 1 cm$^2$. The inner surface was  biased, either to +30 V or to -30 V, to measure the electron or ion saturation current, respectively. The outer surface of the 2FP was not biased, and therefore the floating potential was measured. The probe was mounted at 10 mm from the racetrack, see Figure \ref{fig1}. To correlate the light emitted from the racetrack and the electric signal measured at the 2FP, the light emitted from the racetrack, closest to the 2FP, was collected using two apertures and photomultiplier tube (PMT). Fast ICCD camera measurements facing the target, with acquisition time of 100 ns was used, as described in \cite{ehiasarian_high_2012}. \\
Figure \ref{fig2} shows fast ICCD camera shots of discharge with Ti (13.6 eV), Nb (14.0 eV), Cr (16.5 eV) and Al (18.8 eV) targets. The values in the brackets correspond to the second ionization potential of the element. The results show that for elements having lower ionization potential than the first Ar ionization potential (15.8 eV), such as Ti and Nb, the emission shape of the spoke is elongated and diffuse. While for elements having higher ionization potential than Ar, such as Cr and Al, the emission shape of the spoke is triangular with a sharp edge. The emission profiles from discharges with Mo (16.2 eV) and Cu (20.3 eV) (not shown here), exhibit also a triangular shape with a sharp edge. \\
Figure \ref{fig3} and \ref{fig4} presents the comparison between the light signal (top), ion saturation current (middle) and floating potential (bottom) oscillations for the discharge with Al and Ti target, respectively. \\
\emph{Al target}: The light signal of the discharge with Al target exhibits a peak intensity followed by a sharp drop (Figure \ref{fig3}) of intensity that corresponds to the triangular shape emission profile shown in Figure \ref{fig2}d. Results presented in Figure \ref{fig3} show a simultaneous increase in the optical emission, in the ion saturation current, and a drop in the floating potential. When applying a positive bias it was observed that peaks of the electron current correspond in time and shape to the occurrence of the peaks in the floating potential. Therefore, a drop in the floating potential can be interpreted as an increase in the electron flux. The signal of ions and electrons at the position of the 2FP provides twofold information. First, that the high plasma emission region can be identified as the ionization zone, being suggested by Anders et al. \cite{anders_drifting_2012}; second, that the charged particles diffuse across the magnetic field lines reaching the 2FP, indicating strong cross field transport of electrons and ions, sometimes referred to as anomalous diffusion \cite{brenning_faster-than-bohm_2009}. The oscillations presented in Figure \ref{fig3} have typical profiles of elements with second ionization potential higher than Ar, such as Al.\\
\emph{Ti target}: The emission profile for a Ti target shows a diffuse shape without any sharp drop in intensity, as shown in Figure \ref{fig4} and Figure \ref{fig2}a. Even though the plasma emission differs, it is possible to recognize the localized production of energetic electrons. As in Figure \ref{fig3}, the drop in floating potential is closely followed by an increase in ion saturation current and plasma emission. However, instead of a sharp drop in ion saturation current and plasma emission the ion current remains high for a certain time and the plasma emission exhibits a smooth decay. We postulate that this different shape is dominated by the different dynamics of the generation of secondary electrons. The oscillations presented in Figure \ref{fig4} have typical profiles for elements with second ionization potential lower than Ar, such as Ti. \\
The results indicate that two predominant phenomena are taking place: anomalous transport of energetic electrons, and different shapes of the plasma emission for different materials.


\section{Anomalous electron transport}

In the ionization region of conventional magnetrons the electron pressure is negligibly small compared to the confining magnetic pressure. 
Under the extreme conditions of a HiPIMS discharge, however, this may not be the case. \linebreak
We calculate the pressure of the highly energetic (or \emph{hot}) electrons localized in the spoke and argue that they possess enough free energy to trigger an anomalous diffusion process.
To that end, we construct an effective kinetic equation for the spoke-averaged hot electron distribution function $F(\eps,t)$. 
The spoke volume is $V= A H$, where $A$ is the area (seen from the top), and $H$ is the height (seen from the side). \linebreak
Further dynamic quantities are the cold electron density $n_{\rm e}^{\rm cold}$, with temperature $T_{\rm e}$, and the ion density $n_{\rm i}$.\linebreak Quasineutrality holds and reads
\begin{equation}
	\int_{\eps_{\rm min}}^{e U_{\rm d}} F(\eps) \ d\eps + n_{\rm e}^{\rm cold} = n_{\rm i}. \label{qn}
\end{equation}
We assume that the hot electrons enter the spoke region with energy $e U_{\rm d}$ at a rate $\gamma \Gamma_{\rm i}$. 
Here, $\gamma$ is the secondary electron emission coefficient and $\Gamma_{\rm i} = n_{\rm i} u_{\rm B} = n_{\rm i} H /\tau_{\rm i}$ is the ion flux that reaches the target with the characteristic Bohm velocity $u_{\rm B}=\sqrt{T_{\rm e}/m_{\rm i}}$. 
Moreover, we have introduced the ion lifetime $\tau_{\rm i} = H/u_{\rm B}$.
Confined by the magnetic field and the target potential, the electrons undergo a bouncing motion until they have lost all their energy by inter\-action with the gas and metal neutrals (and emerge as \emph{cold}),\linebreak
or until they have diffused out of the ionization region.  
\linebreak
With respect to the ionization channel, we assume that the newly generated electrons are cold,
and introduce an effective collision frequency $\nu_{\rm ion}$ and an effective ionization energy $E_{\rm c}$, i.e. the frequency at which electron-ion pairs are generated, and the average dissipated energy respectively.\linebreak For argon, $E_{\rm c} \approx 25\,{\rm eV}$ at high electron energies \cite{2005-LiebermanLichtenberg}.\linebreak
We set $\nu_{\rm ion}  = e U_{\rm d}/(E_{\rm c} \tau_{\rm v})$, where $\tau_{\rm v}$ is the average life\-time of the hot electrons with respect to the volume losses, i.e. inelastic collision processes.\linebreak
For the diffusion losses, we define a loss frequency $1/\tau_{\rm d}$.\linebreak
These considerations allow us to formulate the kinetic equation for the hot electrons as
\begin{equation}
	\frac{\dd F}{\dd t} = \frac{e U_{\rm d}}{\tau_{\rm v}} \frac{\dd F}{\dd \eps} - \frac{1}{\tau_{\rm d}}F. \label{kineq}
\end{equation}
The boundary condition at $\eps = e U_{\rm d}$ expresses the assumption that the total influx $\gamma \Gamma_{\rm i} A$ of hot electrons is distributed over the spoke volume $V$: 
\begin{equation}
	F(e  U_{\rm d}) = \frac{\gamma \Gamma_{\rm i} A }{E_{\rm c} \nu_{\rm ion} V}  =
	\gamma \frac{\tau_{\rm v}}{\tau_{\rm i}} \frac{ n_{\rm i}}{ e U_{\rm d}}. \label{bbc}
\end{equation}
The balance equation for the ions accounts for volume generation by ionization and surface losses by diffusion to the wall,
\begin{equation}
\frac{\partial n_{\rm i}}{\partial t} = \frac{1}{\tau_{\rm v}} \frac{ e U_{\rm d}}{E_{\rm c} } \int_{\eps_{\rm min}}^{e U_{\rm d} } F(\eps)\, d\eps - \frac{1}{\tau_{\rm i}}\, n_{\rm i}, \label{iondyn}
\end{equation}
where $\eps_{\rm min}$ is the minimal hot electron energy, i.e. about 10$\% \ e U_{\rm d}$.\\
Thus, we obtain a set of coupled differential equations, governed by the three time constants $\tau_{\rm i}$, $\tau_{\rm v}$, and $\tau_{\rm d}$.\linebreak The system can be reduced  by  observing that the PDE (\ref{kineq}) is solved by any function of the type \linebreak $g(t/\tau_{\rm v} + \eps/(e U_{\rm d})) \cdot \exp(\tau_{\rm v} \eps/(e \tau_{\rm d} U_{\rm d}))$; the boundary condition (\ref{bbc}) then gives the distribution 
\begin{equation}\hspace{-0.04cm}
F(\eps,t)\!=\!
\frac{  \tau_{\rm v}  }{\tau_{\rm i} }\!  \frac{\gamma}{e U_{\rm d}}\! \exp\!\left(\!{\frac{\tau_{\rm v} (\eps -e U_{\rm d})}{ \tau_{\rm d}e U_{\rm d}}}\!\right) \!
n_{\rm i}\!\!\left(\!t+\frac{\tau_{\rm v}\eps}{e U_{\rm d}}-\tau_{\rm v}\!\right)\!.
\end{equation}
Inserting this into the ion balance equation (\ref{iondyn}) yields a single integro-differential equation for the ion density
\begin{equation}
\tau_{\rm i}\frac{\partial n_{\rm i}}{\partial t} = \gamma\frac{eU_{\rm d}}{ E_{\rm c}} \int_0^{1}\!     \exp\!\left(\!{-\eta\frac{\tau_{\rm v} }{\tau_{\rm d}}}\!\right)
 \!  n_{\rm i}\left(t-\eta\,\tau_{\rm v}   \right)  d\eta - n_{\rm i}(t).
\end{equation}
This equation is linear and homogeneous, and its most natural solution is an exponential. 
Employing the ansatz $n_{\rm i}(t) = \hat{ n} \exp \left(t /\tau \right)$ and defining the effective multi\-plication factor
$ m = \gamma e U_{\rm d}/E_{\rm c}$, we obtain a nonlinear\linebreak  equation for the time constant $\tau$:
\begin{equation}
1 - m\frac{ 1-\exp\left(-   \left(\tau_{\rm v}/\tau+\tau_{\rm v}/\tau_{\rm d}\right)\right)}
	{ \tau_{\rm v}/\tau+\tau_{\rm v}/\tau_{\rm d}} + \frac{\tau_{\rm i}}{\tau}  = 0.
\end{equation}
Figure \ref{fig3} shows that the ion current exhibits an exponential growth before the occurrence of a pronounced electron transport, and a steady state as the anomalous transport occurs. We examine the case before the occurrence of the anomalous transport, setting the diffusion constant $\tau_{\rm d}\longrightarrow\infty$, and the steady state case for the ion density, setting $\tau\longrightarrow\infty$. In the latter case, the diffusion term (or hot electron transport) is taken into account. Setting $H = 1$ cm, $\gamma = 0.1$, $T_{\rm e} = 5$ eV and $U_{\rm d} = 500$ V, we estimate the volume loss constant to be $\tau_{\rm v}= 1\,\mu{\rm s}$, and the ion time constant to be $\tau_{\rm i}= 0.3\,\mu{\rm s}$.\\
During the early phase, when no hot electron diffusion takes place, we indeed obtain a fast exponential growth, with $\tau\approx 1.8\,\mu{\rm s}$.  \\
During the second phase we obtain a diffusion loss constant $\tau_{\rm d}\approx 0.6\,\mu{\rm s}$, representing the time scale for locally generated hot electrons to diffuse out of the closed magnetic field region, which corresponds the time scale of the “jet” reported by Ni et al. \cite{ni_plasma_2012}. The calculated density of hot electrons is $n_{\rm e}^{\rm hot}\approx 0.36 \ n_{\rm i}$, and the average energy of hot electrons calculated as
\begin{equation}
	\langle \eps^{\rm hot}\rangle = \dfrac{1}{n_{\rm e}^{\rm hot}} \int_{\eps_{\rm min}}^{e U_{\rm d}} F \ \eps \ d\eps
\end{equation}
results $\langle \eps^{\rm hot}\rangle \approx 200$ eV. Therefore, in order to estimate the hot electrons pressure, it is necessary to estimate the ion density. 
We set $A = 1$ cm$^2$, $I_{\rm d} = 100$ A, and we assume that half of the discharge current is driven through the spoke. Given the two-species nature of the sheath in this late phase, the Bohm velocity is defined by employing the average energy of the hot electrons. Under these assumptions, the ion density results $n_{\rm i} = I_{\rm d}/ \left(2 e A \sqrt{ \left(2/3\right) \langle \eps^{\rm hot}\rangle/m_{\rm i}}  \right) \approx 1.75 \cdot 10^{20}$ m$^{-3}$, which is a reasonable value at this gas pressure. \\
Finally, the hot electron pressure can be estimated as $p^{\rm hot} =  n_{\rm e}^{\rm hot} \left(2/3\right)\ \langle \eps^{\rm hot}\rangle  \approx 1.34$ kPa. On the other hand, the average magnetic pressure in the ionization region, with $B \approx 50$ mT, results $p_{\rm B} = B^2/(2 \mu_0) \approx 1$ kPa.\\
These rough estimates lead us to the conclusion that in the high emissivity region, where most of the secondary electrons are produced and most of the inelastic collision processes take place, $p^{\rm hot}$ exceeds $p_{\rm B}$. Therefore \emph{locally} the high hot electron pressure represents a reservoir of free energy, that the magnetic pressure cannot balance. If we postulate this free energy to give rise to local electron deconfinement and enhanced cross field diffusion, then the streak camera images reported by Ni et al. \cite{ni_plasma_2012} can be explained: hot electrons generated at the target, diffusing across the magnetic field lines and leaving the magnetic confinement region (still obeying quasineutrality) excite the atoms and ions on their path causing the observed trail of emission.
\section{Explanation of sharp emission cutoff}
We postulate the sharp emission cutoff to be a consequence of gas rarefaction and suppressed secondary electrons production. Localized sputtering processes will yield a large number of target atoms. These atoms are ionized and accelerated back to the target, sputtering the target material and generating additional secondary electrons. The transit time over a point on the racetrack of a spoke is about one $\mu$s, as shown in Figure \ref{fig3}. The time needed to an Ar or Al ion to be accelerated from distance of 1 mm to the target by an electric field of $\approx$ 104 V/m \cite{rauch_plasma_2012} is around 250 ns. Therefore, several sputtering cycles can be expected to take place within a single spoke. Assuming half of the discharge current to be distributed only in the spoke, it is reasonable to expect that abundant sputtering would result in the depletion of Ar neutrals in the target vicinity, i.e. Ar gas rarefaction by sputtering wind 
\cite{kadlec_simulation_2007}, \cite{hecimovic_temporal_2012}. 
The gas rarefaction will locally change the composition of the impinging flux from an Ar dominated to an Al dominated flux, which in turn will hinder the secondary electron production. Energy conservation principles prevent singly charged Al ions from generating secondary electrons: the ionization energy of an Al atom (5.98 eV) should be bigger than two times the electron work function of the sputtering target (4.16 eV), which is not the case \cite{haynes_crc_2012}. This means that only secondary electrons generated during the Ar dominated impinging flux at the beginning of the spoke contribute to ionization and excitation of particles in that same spoke. Excitation and resulting emission will diminish once the energetic electrons reach the open magnetic field lines. Transition to the open magnetic field lines and lack of additional energetic electrons due to an Al dominated impinging flux results in the sharp edge observed in the emission contour (Figure \ref{fig2}d). The already mentioned results with the streak camera reported in ref. \cite{ni_plasma_2012} agree well with the postulate that localized gas rarefaction, and inability of singly charged metal ions to generate secondary electrons, result in a sharp emission edge. 
\section{Plasma emission shapes}
Figure \ref{fig2} shows different shapes of the spoke. We argue that this characteristics can be explained based on the different ionization potentials. Even though singly charged metal ions cannot generate secondary electrons, doubly charged ions have sufficiently high ionization energy to generate secondary electrons. If the doubly charged ionization potential is low enough, then there is a good probability to generate a sufficient amount of doubly charged ions that will contribute in production of secondary electrons. However, the secondary electron generation would continue during the transit of the ionization region with reduced efficiency: indeed the energy required to produce a double charged ion is the sum of the first and the second ionization potentials, and these double charged ions present a reduced secondary electron yield. For instance, $\gamma$  is 0.08 for Ti$^{2+}$ impinging on a Ti target, compared to 0.12 for Ar$^{+}$. These secondary electrons will excite and ionize the species resulting in a continuous emission (i.e. without forming a sharp edge), but with diminishing intensity due to the already mentioned reduced secondary electron generation efficiency.
To address this hypothesis, we measured the plasma emission in a single spoke condition, using a fast ICCD camera for 6 elements with different second ionization potentials as shown in Figure \ref{fig2}. The results show that the emission of elements having lower ionization potential then Ar has a diffuse shape, while the emission of elements having higher ionization potential then Ar have a triangular shape with a sharp edge. These results agree reasonably with the hypothesis that low second ionization potential would result in a diffusely shaped emission, due to the prolonged secondary electron production.\\
In this contribution we postulate a model to describe the observed light inhomogeneities in the HiPIMS discharge, based on the localization of the discharge current. This results in a localized secondary electron generation, which in turn leads to an anomalous diffusion across the magnetic field lines. The anomalous cross field diffusion results from the electron deconfinement due to the pressure of energetic electrons exceeding the magnetic pressure. 
The localization of the ion production and the anomalous electron diffusion has been detected using a double flat probe correlated with a light emission signal. Consequences of the current localization are gas rarefaction and anomalous electron diffusion. These in turn result in a triangular shaped light emission with sharp cut off, for elements with second ionization potential higher than the Ar ionization potential; and in a diffuse shaped light emission, for elements with second ionization potential lower than the Ar ionization potential.\\

This work has been supported by the German Science Foundation (DFG) within the frame of the special research unit SFB-TR 87. The authors gratefully acknowledge fruitful discussions with A von Keudell, T de los Arcos, and  T Mussenbrock.

\end{document}